**Ю. В. Єчкало,**

кандидат педагогічних наук, доцент кафедри фундаментальних і соціально-гуманітарних дисциплін ДВНЗ «Криворізький національний університет»

# ЗАСОБИ НАВЧАННЯ КОМП'ЮТЕРНОГО МОДЕЛЮВАННЯ В КУРСІ ФІЗИКИ

*У статті розглядається проблема розвитку інтелекту студентів у процесі навчання фізики засобами комп'ютерного моделювання. Показано, що засоби комп'ютерного моделювання з фізики є однією зі складових інтелектуально-насиченого навчального середовища. Наведена класифікація моделюючих програмних засобів (демонстраційно-моделюючі програмні засоби та педагогічні програмні засоби типу діяльнісного предметно-орієнтованого середовища). Доведено, що адекватне застосування педагогічних програмних засобів типу діяльнісного предметно-орієнтованого середовища в курсі фізики сприятиме розвитку інтелекту студентів.*

***Ключові слова****: курс фізики, засоби комп'ютерного моделювання, розвиток інтелекту.*

*Постановка проблеми у загальному вигляді та її зв'язок з важливими науковими та практичними завданнями.* Фізика є фундаментальною наукою, яка вивчає загальні закономірності перебігу природних явищ, закладає основи світорозуміння на різних рівнях пізнання природи й дає загальне обґрунтування природничо-наукової картини світу. Фундаментальний характер фізичного знання як філософії науки і методології природознавства, теоретичної основи сучасної техніки і виробничих технологій визначає освітнє, світоглядне та виховне значення курсу фізики як навчальної дисципліни.

Розуміння технології навчання фізики як процесуального способу досягнення цілей навчання на основі узгодженого поєднання організаційних форм, методів і засобів навчання дає підстави виділити інформаційно-комунікаційні технології (ІКТ) навчання. ІКТ стають компонентом змісту навчання фізики, засобом оптимізації та підвищення ефективності навчального процесу, а також сприяє реалізації багатьох принципів розвиваючого навчання.

Одним із найбільш перспективних напрямів використання ІКТ у фізичній освіті є комп'ютерне моделювання фізичних процесів і явищ. Комп'ютерні моделі дозволяють викладачу організувати інноваційні види навчальної діяльності (навчання через дослідження, телекомунікаційні проекти). За умови адекватного використання комп'ютерних моделей можна вирішити багато завдань навчання фізики та розвитку особистості, зокрема: формування у студентів системи фізичного знання на основі теоретичних моделей; оволодіння студентами методологією природничо-наукового пізнання і науковим стилем мислення; формування у студентів загальних методів та алгоритмів розв'язування фізичних задач, евристичних прийомів пошуку розв'язку проблем адекватними засобами фізики; розвиток у студентів узагальненого експериментального вміння вести природничо-наукові дослідження методами фізичного пізнання; формування наукового світогляду студентів [2].

*Аналіз останніх досліджень та публікацій, в яких започатковано розв'язання даної проблеми і на які спирається автор.* Окремі напрямки використання комп'ютерного моделювання у навчальному процесі вищого навчального закладу досліджені в ряді робіт вітчизняних фахівців з методики навчання фізики: В. Ф. Заболотного, О. І. Іваницького, Л. Р. Калапуші, М. І. Садового, О. М. Соколюк, Н. Л. Сосницької, І. О. Теплицького.

*Виділення невирішених раніше частин загальної проблеми, яким присвячується означена стаття.* Огляд літератури з теорії та методики навчання фізики свідчить, що проблема розвитку інтелекту студентів засобами комп'ютерного моделювання є недостатньо дослідженою.

*Формулювання цілей статті (постановка завдання).* Виявити засоби комп'ютерного моделювання, застосування яких в курсі фізики сприятиме розвитку інтелекту студентів.

*Виклад основного матеріалу дослідження.* Навчальне середовище можна визначити як штучно побудовану систему, структура і складові якої створюють необхідні умови для досягнення цілей навчально-виховного процесу. Однією зі складових навчального середовища є система засобів навчання як сукупність матеріальних та інформаційних об'єктів, що можуть застосовуватися студентами та викладачами протягом навчання і в яких задовольняються вимоги щодо їх ефективного та безпечного використання [1].

М. Л. Смульсон визначає, зокрема, такі характеристики інтелектуально-насиченого середовища, яке проектується як комплексний навчальний вплив (під характеристиками середовища автор розуміє, зокрема, ті параметри учбової діяльності, які воно забезпечує):

– проблемність і невизначеність середовища (континуум проблемних ситуацій проектується як такий, що конституює середовище, при цьому відбувається самостійне бачення проблемних ситуацій в ньому, самостійність при переході до розв'язування задач, самостійна постановка і розв'язування задач, має місце багатоваріантність засобів розв'язування, можливих рішень і критеріїв розв'язку, багатоваріантність ментальних репрезентацій задач);

– надпредметність середовища (використання надпредметного змісту інтелекту інтелектуальної діяльності, в тому числі винахідницьких проблем, а також проблем з елементами соціального та екзистенціального змісту);

– метакогнітивний характер середовища (інтелектуальна діяльність у середовищі супроводжується метакогнітивним, рефлексивним її моніторингом, усвідомленням характеристик середовища, особливостей групової діяльності в ній, усвідомленням структури і функцій інтелекту, механізмів інтелектуального розвитку в середовищі);

– процесуальність середовища (середовище має єдину часово-просторову структуру, в якій є очевидною цінність процесу інтелектуальної діяльності, а не тільки і не стільки її результату);

– інтелектуальна потенційність середовища (воно постійно висвітлює, «викриває» всі аспекти діяльності учасників, які свідчать про наявність у них інтелектуального потенціалу та розвиток інтелекту);

– інтегративно-діяльнісний характер середовища (ті складники інтелекту, які формуються, постійно інтегруються, «зливаються», «створюють коаліції» в інтелектуальній діяльності);

– децентрованість середовища (потенційна наявність більше ніж одного варіанта майже кожного кроку як індивіда, так і всієї групи в середовищі);

– груповий характер середовища. Це означає, що, по-перше, середовище проектується як таке, у якому відбувається групова (спільна) інтелектуальна діяльність в умовах довіри, взаєморозуміння, безпеки, по-друге, в її активізації використовуються ефекти групової динаміки [11].

Засоби навчання – матеріальні й ідеальні об'єкти, що використовуються в освітньому процесі як носії відомостей (інформаційних ресурсів) та інструменти діяльності викладача та студентів, що застосовуються ними як окремо, так і спільно [12, с. 230].

До засобів навчання належать: природне і соціальне оточення, обладнання, підручники, книги, наукові видання, комп'ютери і комп'ютерні мережі з відповідним програмним забезпеченням та інформаційними ресурсами, зокрема електронні підручники, довідники, енциклопедії, електронні бібліотеки [6].

Існують різні класифікації засобів навчання. Одна з них – класифікація за дидактичною функцією [9, с. 158]:

– інформаційні засоби (підручники, навчальні посібники та ін.);

– дидактичні засоби (таблиці, плакати, відеофільми, програмні засоби навчального призначення, демонстраційні приклади та ін.);

– технічні засоби навчання (аудіовізуальні засоби, комп'ютери, засоби телекомунікації, системи мультимедіа, віртуальна реальність та ін.).

Людська пам'ять найбільш ефективно зберігає інформацію при сполученні роботи зорового і слухового каналів її одержання, тому особливе місце серед технічних засобів навчання посідають системи мультимедіа.

Мультимедіа являє собою засіб, за допомогою якого реалізуються ідеї інтенсифікації навчання, спрямовані на пошук максимально ефективних методів і засобів навчання, адекватних його цілям і змісту; інтеграції педагогічної науки, практики; цілісності і безперервності педагогічного процесу.

Дидактичне призначення використовуваних програмних засобів може бути різним: опанування нового матеріалу (наприклад, за допомогою програми навчального призначення), закріплення нового матеріалу (наприклад, за допомогою програми-тренажера), перевірка рівня засвоєння навчального матеріалу або операційних навичок (наприклад, за допомогою програм автоматизованого контролю або тестування).

За Ю. І. Машбицем, ІКТ у сучасному навчальному процесі використовуються як [7; 8]:

– предмет навчання (у ході вивчення конструкції та принципу дії апаратних засобів, алгоритмічних і машинних мов та ін.);

– засіб навчання (шляхом використання відповідним чином сконструйованих «навчальних програм», тобто програм прямої навчально-формуючої дії);

– засіб навчальної діяльності (за допомогою спеціально розроблених програм підтримки навчальної діяльності у даній предметній галузі або у ході використання прикладного програмного забезпечення загального призначення).

Відмінність між другим та третім напрямками полягає у способі використання засобів ІКТ, який зумовлює шляхи організації навчально-пізнавальної діяльності студентів. У рамках другого підходу процес навчання потрібно розглядати з точки зору управління навчальною діяльністю, основним засобом якого є навчальний вплив на пізнавальну сферу студентів.

Тому проектування навчальних програм будь-якого типу повинно базуватися на наступному психолого-педагогічному підґрунті [8]:

– розширення можливостей подання навчальної інформації;

– підсилення мотивація учіння;

– зростання активності діяльності студентів;

– розширення наборів задач, що застосовуються у навчанні;

– якісна зміна контролю за діяльністю студентів та забезпечення гнучкості управління навчальним процесом;

– сприяння формуванню у студентів рефлексії своєї діяльності.

Відомо, що такі засоби навчання, як персональний комп'ютер викладача та персональні комп'ютери студентів, а також комп'ютерні програми навчального призначення формують матеріальну складову навчального середовища та беруть участь у навчальній діяльності. Разом із ІКТ навчання вони мають функцію засобів діяльності учасників навчального процесу – студента і педагога. Сучасні ІКТ надають студентам широкі можливості при проведенні експериментів, виявленні різноманітних проявів закономірностей, що мають місце у природі та техніці, систематизації спостережуваних фактів, створюючи тим самим умови для активізації розумової діяльності. Комп'ютерне моделювання фізичних процесів і явищ є одним з найбільш перспективних напрямів використання ІКТ у фізичній освіті.

Комп'ютерне моделювання в курсі фізики зіштовхується з рядом труднощів суто технічного плану – програмування. На створення комп'ютерної моделі «з нуля» потрібен досить великий час, що навряд чи можливо в рамках одного-двох занять. Крім того, велика частина роботи виявляється спрямованою на програмування сервісних, другорядних операцій. Сама модель звичайно полягає в організації нескладного алгоритму. Ці причини і є основними перешкодами на шляху використання комп'ютерного моделювання в навчанні фізики. Але всього цього можна уникнути, якщо звільнити студента, що займається комп'ютерним моделюванням, від рутинної програмістської роботи.

На сьогодні створено багато педагогічних програмних засобів (ППЗ), використання яких у навчальному процесі з фізики поряд із традиційними засобами діяльності сприяє поліпшенню якості навчання, підвищенню рівня теоретичних знань та практичних вмінь та навичок студентів, активізує навчально-пізнавальну діяльність тощо.

Аналіз літературних джерел показує, що зараз немає єдиної класифікації моделюючих програмних засобів. М. І. Жалдак, В. В. Лапінський, М. І. Шут виділяють демонстраційно-моделюючі програмні засоби та ППЗ типу діяльнісного предметно-орієнтованого середовища [3].

Характерними ознаками демонстраційно-моделюючих програмних засо-бів є їх використання на етапах пояснення нового матеріалу, фронтальної демонстрації моделі об'єкту вивчення. Можливі варіанти ППЗ, які відрізняються за способом формування моделі, видом моделі. Можна виділити:

а) імітаційні неінтерактивні моделі, які виконують роль динамічних плакатів;

б) імітаційні інтерактивні моделі, характерним для яких є зовнішня схожість з об'єктом вивчення (фізичним явищем, природнім об'єктом тощо), яка формується з використанням математичної моделі, суттєво відмінної від тієї, яка використовується для наукового опису цього явища, тому математичний опис моделі є закритим для студента;

в) інтерактивні моделі, засновані на математичних описах явищ, максимально наближених до наукових моделей певної предметної галузі і тому відкритих (або частково відкритих, доступних) для студента [3].

Більшість викладачів та методистів (зокрема, автори [4; 5]) надає перевагу роботі з готовими комп'ютерними моделями. Такий вид організації навчальної діяльності має очевидні переваги – можливість реалізації дидактичного принципу наочності, встановлення міжпредметних зв'язків, підвищення якості знань, створення позитивної мотивації, посилення інтересу студентів до предмету. Однак разом з цим він має і суттєві недоліки: оскільки не розкриваються механізми створення моделей, робота студентів з ними носить переважно відтворюючий характер.

До ППЗ типу діяльнісного предметно-орієнтованого середовища можна віднести моделюючі програмні засоби, призначені для візуалізації об'єктів вивчення та виконання певних дій над ними. Засоби цього типу іноді називають «мікросвітами» та відносять до інтелектуальних навчальних систем, в яких має місце научіння шляхом так званого «занурення» студента в середовище.

Мікросвіти по суті є моделями реального світу. Висувається ряд критеріїв для побудови мікросвіту фізики [10]:

1) вивчення руху в роботі з простими для студента законами руху;

2) можливість дій, ігор та ін., які були б справжньою діяльністю в цьому

мікросвіті;

3) мікросвіт повинен бути побудований так, щоб всі необхідні поняття могли б бути визначені на основі роботи в цьому мікросвіті.

У такому мікросвіті фізики студент повинен мати можливість висувати власні припущення про нього і його закони, перевіряти їх правильність, будувати модель, з якою буде працювати, перевіряти її на адекватність або запропонувати шляхи її покращення. У такий спосіб конструюється навчальне середовище, в якому студент стає активним проектувальником власного учіння.

Правила в діяльнісних навчальних середовищах складає розробник системи. Потім вони можуть бути розвинені, змінені або доповнені самими студентами. У проблемі мікросвітів принципово важливим є те, що в такому інтелектуальному навчальному середовищі критерії «вірно-невірно» не посідають «домінуючого положення». Традиційно помилка, невірна відповідь означають, що студент не слухав, не вчив або не може нічого зрозуміти. Однак у роботі з мікросвітами кожна помилкова дія виступає як джерело нових творчих ідей, крок до одержання творчого і особистісно значущого результату, розвитку інтелекту.

Хоча функції навчання як управління учбовою діяльністю в мікросвітах зведені до мінімуму, однак вони можуть давати значний педагогічний ефект завдяки тому, що тут можливості інтелектуальних навчальних систем спрямовані на активізацію творчої складової діяльності, позитивного емоційного фону учіння.

Також до цього типу ППЗ можна віднести різного виду тренажери, симулятори (імітатори). Суттєвою особливістю цього типу ППЗ є їх пристосованість до індивідуального використання студентами. Ці засоби можуть бути призначені для використання як на заняттях, так і в позааудиторній роботі [3].

*Висновки за результатами дослідження.* Основними завданнями навчання комп'ютерного моделювання в курсі фізики є загальний розвиток і становлення світогляду студентів, оволодіння моделюванням як методом пізнання, вироблення практичних навичок комп'ютерного моделювання, реалізація між-предметних зв'язків, розвиток і професіоналізація навичок роботи з комп'ютером, формування навичок проектної діяльності.

До засобів навчання комп'ютерного моделювання відносяться демонстраційно-моделюючі програмні засоби та педагогічні програмні засоби типу діяльнісного предметно-орієнтованого середовища, застосування яких сприяє глибокому розумінню суті логічних відношень між оригіналом і моделлю, особливостей побудови моделей, формуванню у студентів уявлення про моделювання як про метод пізнання навколишнього світу.

Надання студентам можливості моделювання фізичних процесів і явищ, організації на цій основі їх експериментально-дослідницької діяльності є методологічною основою роботи викладача з розвитку інтелекту студентів.

**Средства обучения компьютерному моделированию в курсе физики**
Ечкало Ю. В.

*В статье рассматривается проблема развития интеллекта студентов в процессе обучения физике средствами компьютерного моделирования. Показано, что средства компьютерного моделирования физических процессов являются одной из составляющих интеллектуально-насыщенной учебной среды. Приведена классификация моделирующих программных средств (демонстрационно-моделирующие программные средства и педагогические программные средства типа деятельностной предметно-ориентированной среды). Доказано, что адекватное использование педагогических программных средств типа деятельностной предметно-ориентированной среды в курсе физики способствовать развитию интеллекта студентов.*

***Ключевые слова**: курс физики, средства компьютерного моделирования, развитие интеллекта.*

**Tools of computer simulation in learning physics**
Yechkalo Yu. V.

*The article deals with the problem of intellectual development of students in*

*learning of physics by means of computer simulation. It is shown that the means of computer simulation of physical processes is one of the components of intellectual learning environment. It is spoken in detail about classification of simulation software such as software of demonstration and modelling and educational software tools which means environment of activity. It is proved that the adequate use of educational software tools which means environment of activity in learning of physics is contribute to the development of intelligence of students.*

***Key words****: learning physics, computer simulation tools, development of intellect.*